\documentclass{INTERSPEECH2023}
\DeclareMathOperator{\e}{e}
\usepackage{relsize}

\interspeechcameraready

\title{Why can big.bi be changed to bi.gbi? A mathematical model of syllabification and articulatory synthesis}
\name{Fr\'ed\'eric Berthommier}
\address{Univ. Grenoble Alpes, CNRS, Grenoble INP, GIPSA-lab, 38000 Grenoble, France}
\email{frederic.berthommier@gipsa-lab.grenoble-inp.fr}

\begin{document}

\maketitle
 
\begin{abstract}
A simplified model of articulatory synthesis involving four stages is presented. The planning of articulatory gestures is based on syllable graphs with arcs and nodes that are implemented in a complex representation. This was first motivated by a reduction in the many-to-one relationship between articulatory parameters and formant space. This allows for consistent trajectory planning and computation of articulation dynamics with coordination and selection operators. The flow of articulatory parameters is derived from these graphs with four equations. Many assertions of Articulatory Phonology have been abandoned. This framework is adapted to synthesis using VLAM (a Maeda's model) and simulations are performed with syllables including main vowels and the plosives /b,d,g/ only. The model is able to describe consonant-vowel coarticulation, articulation of consonant clusters, and verbal transformations are seen as transitions of the syllable graph structure.
\end{abstract}

\section{Introduction}
Articulatory synthesis is a tool for understanding phonological processes because it involves the definition of articulatory gestures followed by the synthesis of real speech sounds. Its goal is to specify the complete pathway between linguistic structures and the physical world of vocal tract movement and acoustics. Articulatory phonology (AP) \cite{Browman1992} combined with task dynamics (TD) \cite{Saltzman1989} is the reference model in this field. The two main steps are the planning of the gestural score from a phonetic string and the conversion of this plan into the movement of an articulatory model. The gestural score includes the activation period of each articulator and the phasing which constitutes the coordination. Despite its rigorous design, TD has only recently been implemented for VCV synthesis \cite{Alexander2019} with similar results as those discussed here. The complexity of this implementation stems from the distinction between the status of vowels and consonants. Vowels are spatial configurations of the vocal tract that are articulator configurations while consonants are dynamically defined with constriction goal variables. On the other hand, it has long been shown by Carré et al. \cite{CARRE2017} that vowel space emerges from the exploitation of an acoustic tube and that simple rules allow for the synthesis of VCV syllables \cite{CARRE1995}. This theory was physically grounded and extended by Story and Bunton \cite{Story2009,Story2019} to a tube model having more sections. This was constructed from the human vocal tract and it has a more precise locus of constriction but we distinguish it because it is controlled section by section to calculate the area function (see \cite{Kroger2022} for a review about articulatory models). The gap between tube and articulatory models is resolved by the current approach in which consonants and vowels have a unified articulatory-acoustic representation. Schwartz et al. \cite{Schwartz2012} have shown with a Maeda's model (named VLAM hereafter) that the plosives /b,d,g/ have their formants localized around the vowel space. The first step in this approach is to establish a bijection between the articulatory space of this model and the vowel space, which allows us to define trajectories between points. Thus a trajectory drawn at this level has an acoustic image which is a modulation of formants. This direct path between planning and acoustic effects greatly simplifies control and practically, the small set of parameters of the model can be adjusted manually by simple hearing. This model opposes other approaches in several ways. First at all, by looking for mathematical relationships as AP/TD does, this approach is the opposite of the large simulations undertaken to find the same type of articulatory trajectories for syllable modeling \cite{vanNiekerk2022,vanNiekerk2023}. It meets the demand for less energy consumption and the mathematical relationships are valuable. By giving a structural point of view, they might constrain much more than any AI approach \cite{dupoux2018} the research on the biological foundations of language. The vocal tract model is minimalist and provides the features just necessary to produce syllables with main vowels and /b,d,g/. It has a jaw unlike the tube models. VLAM formants are obtained with a classical transmission line method \cite{BadinFant1984}. The source model is also rudimentary and the only post-processing applied at the output of the synthesis is a multiplication by an envelope depending on the syllable structure. This framework is supposed to be adaptable to different VT geometries and other phonemes. Coarticulation mechanisms are revisited and separability between articulators assigned to vowel trajectories and those assigned to consonant trajectories has been found as with the tube models in the line opened by \cite{Ohman1966,Ohman1967}. We propose to plan syllable and word production in a bottom-up manner using graphs and interpret verbal transformations as transformations of these graphs. These contain all the information for making the selection and coordination of the 7 VLAM articulators (see Tilsen's discussion \cite{Tilsen2013} on the concepts of selection and coordination in the context of AP). The syllabic segmentation is not well understood and it is a hot topic currently investigated with DNNs \cite{Liu2023}.To understand syllabification, the present paper offers a new generative formalism which reflects production processes without much reference to their biological implementation (as \cite{Guenther2006}) but physically well grounded (as \cite{Story2009,Story2019}).
\section{Description of the model}
\begin{figure}[t]
  \centering
  \includegraphics[width=\linewidth]{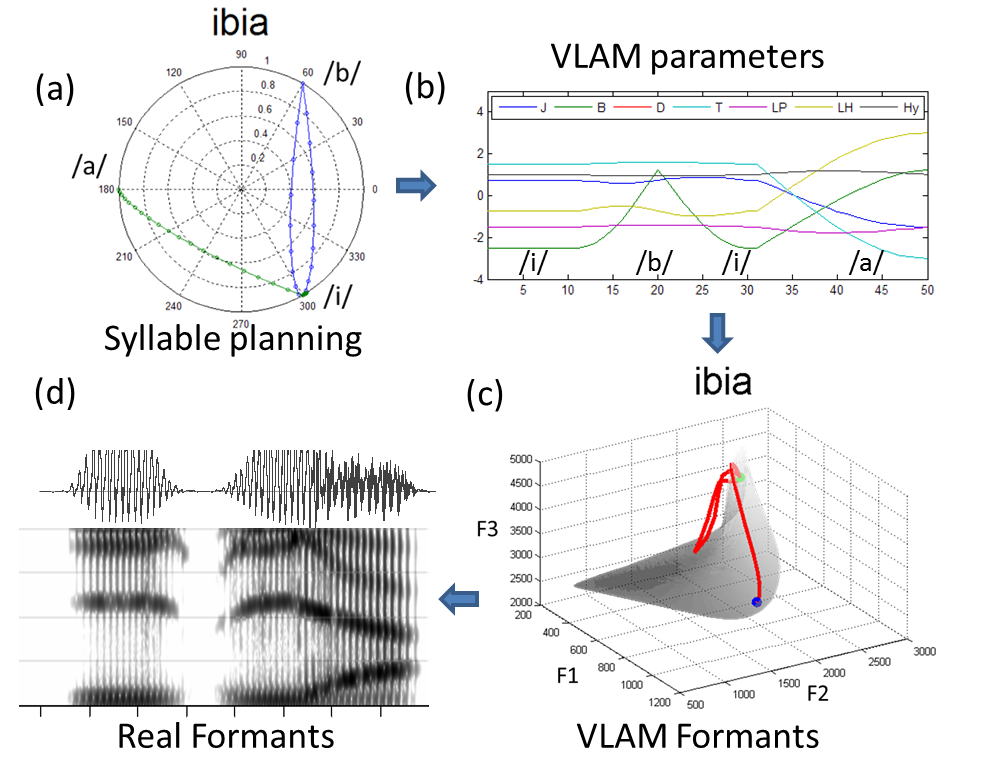}
  \caption{The four steps of synthesis of /ibia/ with $T=100 ms$. (a) Planning trajectories $z_v(t)$ and $z_c(t)$. (b) Flow of articulatory parameters showing the trough effect (variation of Body during /b/). (c) Formant trajectory in F1-F2-F3 leaving the surface defined by Eq.~\ref{equ:eq1} except during /ia/. (d) Output Spectrogram and wave plot.} 
  \label{fig:Figure1}
\end{figure}\
\subsection{Equations of the model}

The model is based on the reduction of the many-to-one relationship between the articulatory space and the acoustic frequency space of the formants. This reduction is enabled by a coordination function that correlates the articulators of a given model (which can be a tubular model like the DRM or a true articulatory model like the VLAM). The principle is to use physically based symmetries of vowel production to control articulator positions. Such symmetries have been demonstrated by Shroeder \cite{Schroeder1967} and Mermelstein \cite{Mermelstein1967} and then systematically used to control vowel production with the DRM \cite{CARRE2017} (which is an 8-tubes model). We propose a generalization of this type of control for true articulatory models via the coordination function which is parameterized to cover a surface in the F1-F2-F3 space \cite{berthommier2021}. This function realizes a bijection between the complex domain and this surface by computing the parameters and synthesizing vowels with the given articulatory model. The setting of the model is based on the average $\Omega$ and the range $\Psi_1$ of each parameter (here given in VLAM) plus a fixed angle $\Psi_2$ (Table~\ref{tab:defpsy}). The value of each articulatory parameter $P_i, i=1..7$ is computed independently for a given point $(\rho_V,\theta_V)$ of the complex domain. The coordination between $P_i$ is provided by the product of the same complex conjugate $\rho_V\e^{-i\theta_V}$ with each fixed complex value $\Psi_i$ of the VLAM:
\begin{align}
 &P_i-\Omega_i =  Re\left[\Psi_i\rho_V\e^{-i\theta_V} \right]   \nonumber \\
 &\bm{P}-\bm{\Omega} =  Re\left[\bm{\Psi}\rho_V\e^{-i\theta_V} \right] = \rho_V\bm{\Psi_1} \, cos(\bm{\Psi_2} - \theta_V) 
 \label{equ:eq1}
\end{align}
Let remark that the coordination function is a simple cosine. Consistently, the $\Psi_2$ angles are set to have the angles of the corner vowels /iau/ at $\theta=\{\frac{5\pi}{3},\pi,\frac{\pi}{3}\}$ and $\rho=1$. This is obtained by assigning a cardinal vowel to each articulator such that $P_i-\Omega_i=\Psi_{1i}$ for this articulator when $\theta=\Psi_{2i}$. To cover the vowel space and display the whole F1-F2-F3 surface, we choose $\rho_V \in \left[0,1\right]$ and $\theta_V \in \left[0,2\pi\right[$. An extension to a crown coding for consonants is made with $\rho_C \in \left[0,1.2\right]$ and $\theta_C \in \left[0,2\pi\right[$. This surface is shown in Fig.~\ref{fig:Figure1}~c.
\begin{table}[b!]
  \caption{Definition of the coordination function parameters for VLAM.}
  \label{tab:defpsy}
  \centering
  \begin{tabular}{lrcc}
    \toprule
    \textbf{Parameter}      & \textbf{ $\Omega$}      &  \textbf{$\Psi_1$}   &  \textbf{$\Psi_2$}         \\
    \midrule
    1-Jaw                    & $0$                 & $-1.5$               &    $\pi$                  \\
    2-Body                   & $0$                 & $-2.5$               &    $\frac{5 \pi}{3}$               \\
    3-Dorsum                 & $0$                 & $3$                  &    $\frac{\pi}{3}$     \\
    4-Tip                    & $0$                 & $-3$                 &    $\pi$           \\
    5-LipP                   & $0$                 & $3$                  &    $\frac{\pi}{3}$        \\
    6-LipH                   & $0.5$               & $2.5$                &    $\pi$       \\
    7-Hy                     & $0$                 & $-2$                 &    $\frac{\pi}{3}$   \\
    \bottomrule
  \end{tabular}
\end{table}

Trajectories for vowels and consonants are planned in the complex plane by forming arcs between 2 points $(\rho_1,\theta_1)$ and $(\rho_2,\theta_2)$ :
\begin{align}
  z(t) = (1-\rho(t)) \, \rho_1 \, \e^{i \theta_1} + \rho(t) \, \rho_2 \, \e^{i (\theta_2 + \frac{\nu}{K} \theta(t))} 
 \label{eq2}
\end{align}
where $t$ varies between $0$ and $nT$, $\rho(t)=cos(\frac{\theta(t)}{2})$ determines the velocity profile and $\nu=\pm1$ and $K$ are shape parameters. When $K$ is large ($K=30$ for vowel arcs and $K=10$ for consonant arcs), the trajectories become straight lines (see Figure~\ref{fig:Figure1}~a) and we recognize Öhman's equation \cite{Ohman1967}. These parameters are adapted to plan formant trajectories and not to reproduce the real dynamics of the articulators as in TD. There are two orientations of these arcs with origin $a$ and arrival $b$ : $nT_1$ with $\theta(t)=\frac{\pi t}{nT}, a=2, b=1$ or $nT_2$ with $\theta(t)=\pi (\frac{t}{nT} - 1), a=1, b=2$. The number of periods $n$ is important to fix the duration of the vowel arcs as equal to that of the consonant branches. It is the number of consonants plus one (see below).

At each time $t$, the set of parameters is coordinated with Eq.~\ref{equ:eq1} but the modulation is obtained piecemeal via the chaining of time periods of duration $nT$ without discontinuity. Within each period, the trajectories of the parameters are the real part of the product of the complex column vector $\Psi$ and the complex line vector $\bar{z}(t)$ giving matrices of dimension $7*nT$ that are concatenated: 
\begin{align}
 &\bm{P}(t) - \bm{\Omega} =  Re\left[ \bm{\Psi}\bar{z}(t) \right] =
  (1-\rho(t)) \, \rho_1 \, \bm{\Psi_1} \, cos(\bm{\Psi_2} - \theta_1) \,   \nonumber \\
& \qquad\qquad\qquad +\rho(t) \, \rho_2 \, \bm{\Psi_1} \, cos(\bm{\Psi_2} - \theta_2 - \frac{\nu}{K} \theta(t)) 
\label{eq3}
\end{align}

The trajectories during the vocalic segments and pauses are defined by the previous equation whereas the superposition of vocalic and consonant trajectories (named superimposed segment) is obtained by adding a selection process splitting the set of 7 components in two parts coordinated separately but synchronized. The coarticulation between vowels and consonants is due to the superposition of 2 branches having the same duration $nT$ together with the same departure and arrival points which are always vowels $(V_o,V_e)$:
\begin{align}
 &\bm{P}(t) - \bm{\Omega} = Re\left[ S_v.\bm{\Psi}\bar{z}_v(t) + S_c.\bm{\Psi}\bar{z}_c(t) \right]
 \label{eq4}
\end{align}
where $S_v$ and $S_c$ are the two exclusive selection vectors composed of zeros and ones for selected articulators with $S_v+S_c=\bm{1}$ (a column vector of 7 ones). These are multiplied with the complex vector $\Psi$ with a Hadamard product. These vectors depend on the consonant(s), thus avoiding any weighting process. Mathematically, the main effect of superposition is to increase the planning dimension from 2 (1 complex number) to 4 (2 complex numbers). At each instant, two sets of articulators are coordinated separately. This increase is observable in the formant space: while the trajectory of a diphthong remains inside the surface defined by the coordination function, the trajectories of the superimposed segments leave it (see Figure~\ref{fig:Figure1}~c).
\subsection{Assembling syllable graphs}
Syllabication is modeled by graphs embedding all the information necessary for the selection and coordination of articulators. These graphs describe nodes and arcs to calculate with the previous equations a continuous planning and then a continuous flow of parameters even in silent periods. This is a kind micro-grammar of the temporal encoding of syllables gestures as proposed by Gafos \cite{Gafos2002}. This is generative and syllable graphs are concatenated in a coherent way to make words which are separated by pauses. The articulatory model is then permanently animated. This is in contrast to AP which uses the event-based concept of activation vs. inactivation. This requires some clock mechanism (the coupling of oscillators in AP) to define beginning and end of periods of activation. In contrast, timing is considered here as an inherent property of the model structure (as proposed by Fowler \cite{Fowler1980}) and whole duration is externally controlled by a single parameter $T$. This determines formant trajectories and the envelope structure which is superimposed can occlude a part of this information. Nevertheless, the envelope adds up temporal information and the lips movements remain visible for guiding the perceptual recovering of produced syllabic structures \cite{Basirat2012,nabe2022}. 
\begin{figure}[t]
  \centering
  \includegraphics[width=\linewidth]{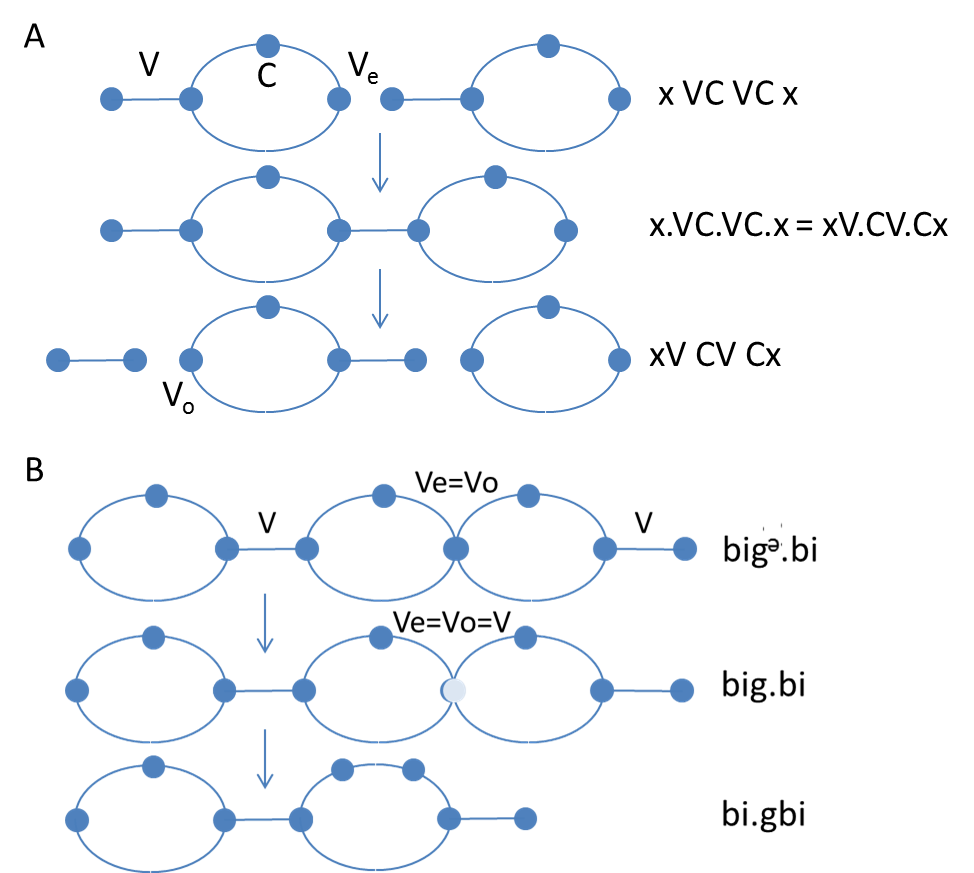}
  \caption{Syllable graphs of two verbal transformations.}
  \label{fig:Figure3}
\end{figure}\

The graph $G=(V,E)$ of a syllable is a concatenation of superimposed segments and purely vocalic segments (e.g. diphthongs). Each superimposed segment has origin and end nodes which are the vowels $V_o$ and $V_e$ (i.e. without selection because all articulators are coordinated together for vowels).  These vowels are not always present in the phonetic chain and must be added as nodes, as we will see, to complete the graph. When they are absent, they can be far from the neutral vowel (with $\rho=0$ and $\bm{P}=\bm{\Omega}$) and they have the same $\theta$ as the closest vowel but $\rho$ is multiplied by specific coefficients $\delta_o$ and $\delta_e$ (which will be set later). Thus, we have $V_o=(\delta_o.\rho_V,\theta_V)$ and $V_e=(\delta_e.\rho_V,\theta_V)$. The value of $\delta_o$ determines the degree of anticipation of visible vowel-related lip movements before the onset of the consonant. 

Let us explicitly define a CV graph. This one has 3 nodes $V=\{V_o,C,V\}$ having a location in the complex plane $L=\{(\delta_o. \rho_V,\theta_V), (\rho_C,\theta_C),(\rho_V,\theta_V)\}$ and 4 arcs $E=\{(V_o,C),(C,V),(V_e,V),(V, V)\}$ having temporal properties $T_p=\{T_1,T_2,2T_2,T\}$ the last one being a stationary point of duration $T$. In this case, $\bm{P}-\bm{\Omega} = Re\left[\bm{\Psi}\rho_V\e^{-i\theta_V} \right]$. These joint properties are necessary to compute the two trajectories $z_v(t)$ and $z_c(t)$ at the planning level. Then, the superposition is applied from the selection properties associated with each arc $S=\{S_c,S_c,S_v,\bm{1}\}$ the last term being neutral (without selection) because this is a stationary vowel. The superposition between vowel and consonant branches is a coproduction mechanism allowing for coarticulation similar to this promoted by \cite{Ohman1966,Ohman1967} (see \cite{Volenec2015} for a review of coarticulation). It removes any weighting of consonant articulation over the vowel as in \cite{Birkholz2013} and its implementation remains as simple as in \cite{CARRE1995}.

For a CVC, the construction is symmetric thanks to the end vowel $V_e$ defined with its free centralization coefficient $\delta_e$. This may depend on the language and it is well known that French tends to produce a schwa-like vocoid attached to the second consonant \cite{Hutin2020}. It is modeled with a low $\delta_e=0.5$ to be salient and with an intermediate $\delta_e=0.7$ to be audible. This is a new important property because the structure of a CVC has now two equivalent superimposed segments. Contrarily to the AP proposal, there is no synchronous vs. sequential implementation of the consonants depending on their onset or coda position. A degree of asymmetry can also be added by varying the period $T$ as a prosodic feature. This feature is flexible, but the superimposed segments need strict synchronization and lengthening could occur only during V before its coarticulation with C. The graph has 5 nodes $V=\{V_o,C_1,V,C_2, V_e\}$ with location $L=\{(\delta_o.\rho_V,\theta_V),(\rho_{C1},\theta_{C1}),(\rho_V,\theta_V),(\rho_{C2},\theta_{C2}),(\delta_e.\rho_V,\theta_V)\}$. The rest is easy to deduce from the CV graph according to the symmetry.

The consonant clusters have a graph relying to the sequencing of several consonants in the same branch. The vocalic anchor point $V_o$ is set as previously with the same anticipation property (e.g. a lips protrusion can appear before the first consonant onset). This timing is different in AP in which the beginning of the vowel is between the two consonants. A CCV has 4 nodes only $V=\{V_o,C_1,C_2,V\}$ with location $L=\{(\delta_o.\rho_V,\theta_V),(\rho_{C1},\theta_{C1}),(\rho_{C2},\theta_{C2}),(\rho_V,\theta_V)\}$ which can be defined jointly for the two consonants (this is another useful feature) and 5 arcs $E=\{(V_o,C_1),(C_1,C_2),(C_2,V),(V_e,V),(V,V)\}$ having timing properties $T_p=\{T_1,T_2,T_2,3T_2,T\}$. The same selection vector is now applied 3 times with $S=\{S_c,S_c,S_c,S_v,\bm{1}\}$ and it is defined specifically for each consonants pair. Despite a similar graph structure, the coarticulation between consonants is more complex because it involves interactions at both levels (location of nodes and selection). 

The concatenation of syllables to make words is achieved by the '.' operator or it is implicit as for V.CV. There are 3 cases $\{xV.Cx, xC.Cx, xC.Vx\}$ to be treated. These are based on the assignation of the anchoring vowels depending on the context: $\{V=V_o,V_e=V_o,V_e=V\}$. Conversely, the pause is defined as the lack of concatenation and coarticulation between two syllables. This is consistently defined as a diphthong transition between the previous $V_e$ and the next $V_o$ with an arbitrary duration $T_p$. These two rules combined have a direct consequence on the syllabification of a succession of syllables VC. There is a transition (which is a verbal transformation) when the time pressure cancels $T_p$ because two consecutive syllables becomes concatenated with $V_e=V$ (the third case seen above with $\delta_e=1$). At this time we have no difference between representations of x.VC.VC.x and xV.CV.Cx because all becomes symmetric ($\delta_o=1$ according the first case above). This does not involve a supplementary temporal organizer as the phase vs. antiphase coupling mechanism of the AP nor an opposition synchronous vs. sequential of onset and coda consonants. This mechanism is also much simpler than that of \cite{Liu2023}. Here, the gain of switching from VC to CV is a cancellation of $T_p$ associated to a verbal transformation represented by graph transitions in Figure~\ref{fig:Figure3}~A. The classical transformation of /ib/ into /bi/ is given in the supplement. 
\section{Simulations}
We give several examples in which the model is successful for representing the Human syllable production. The goal is not to reliably estimate the articulatory parameters because the generated trajectories aim to straightly produce formant modulations and output speech. Moreover, playing with the shape parameters $\nu$ and $K$ indicates there is a great flexibility of trajectory shapes. The dynamics (determined by $\rho(t)$ in Eq.~\ref{eq3}) is somewhat more important but the simulations showed that the key for having correct consonant percepts is the location of nodes in the complex plane combined with the choice of the selection vectors. The model has been tuned for the consonants /b,d,g/ and we found that /b/ is easy to reach with $(\rho_b=1,\theta_b=\pi/3)$ and a selection vector $S_c=(1,1,0,0,0,1,0)^{\mathsmaller T}$ noted with the indices of ones $S_c=\{1,2,6\}$ hereafter. This means that not only lips and jaw are involved for /b/ but that the tongue body parameter is also engaged (see Fig.~\ref{fig:Figure1}~b). The tongue body is making an unexpected front/back movement during /b/ which corresponds to the trough effect \cite{Lindblom2002}. In the continuation of the previous explanation about the verbal transformation, when /i.bi/ is pronounced after the cancellation of $T_p$, the tongue does not stop to move through the consonant. In simulation, the tongue movement as well as the acoustic traces are due to the fact that the placement of the /b/ in the complex plane is the same as for the /u/ vowel. When the selection process is applied, the tongue body follows the $z_c(t)$ trajectory towards /u/ while the vowel articulators $S_v=\{3,4,5,7\}$ are driven by a fairly constant $z_v(t)$ around /i/. The frequency domain trajectory is composite and this is the reason why the locus of /b/ is shifted towards /i/. 
\begin{figure}[t]
  \centering
  \includegraphics[width=\linewidth]{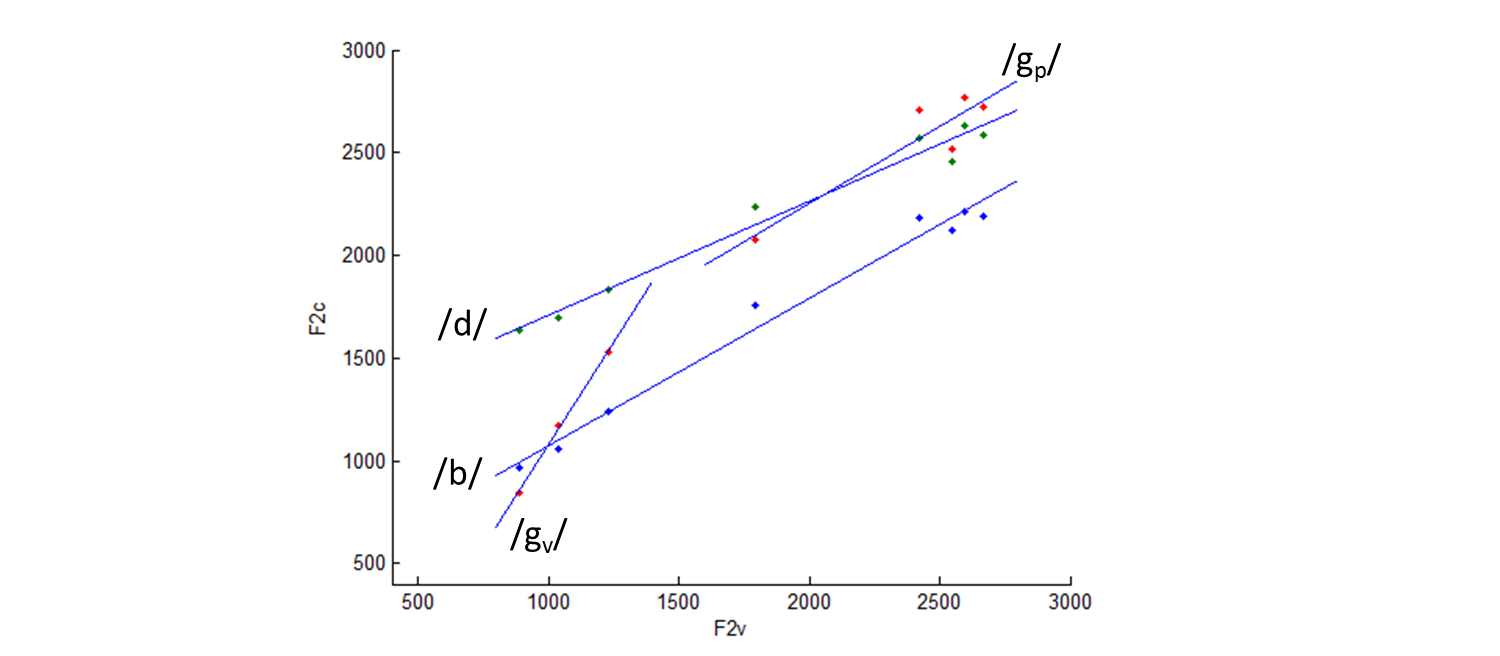}
  \caption{Plot of modelled /b,d,g/ locus equations across the 8 main vowels to be compared to the prototypical locus equations derived in \cite{Sussman1991}.}
  \label{fig:Figure2}
\end{figure}

This raises the question of the degree of compatibility of the model with the locus equations \cite{Sussman1991}. The tuning of /b,d,g/ has been realised by hearing at first and this led to find the conditions of coarticulation of /d/ and /g/ with all vowels. For the /g/ this is well known that it is velar for back vowels and palatal for front vowels. This led to the locations $(\rho_{g_v}=1.2,\theta_{g_v}=\pi/3)$ and  $(\rho_{g_p}=1.1,\theta_{g_p}=23\pi/12)$  combined with the selection vector $S_c=\{1,2,3,4\}$ involving tongue and jaw. The location of /d/ depends on the following vowel around  $(\rho_d=1.2,\theta_d=3\pi/2)$ and /d/ has same selection vector $S_c=\{1,2,3,4\}$. The Figure~\ref{fig:Figure2} is constructed by taking the F2 values 30 ms after the onset for the 8 vowels and $\delta_o=0.5$. This is similar to the classical observation but with a downward shift of the velar /g/ which could be due to geometric differences between the VLAM and the human vocal tract.

\begin{figure}[t]
  \centering
  \includegraphics[width=\linewidth]{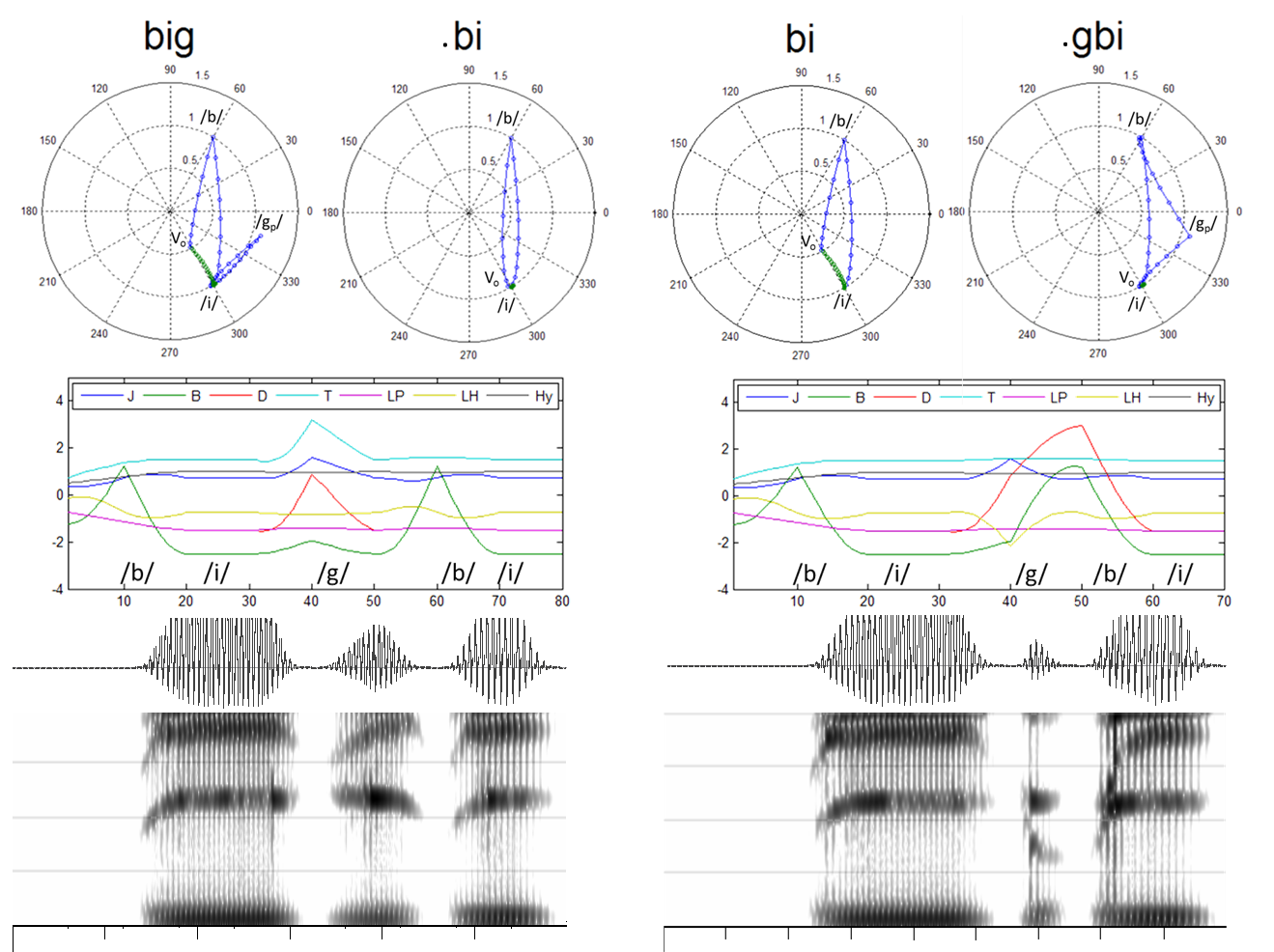}
  \caption{Synthesis of big.bi (with $\delta_o=\delta_e=1$) and bi.gbi. Top: Planning trajectories. Middle: Flow of articulatory parameters. Bottom: Aligned wave plots and output spectrograms. The envelope is similar after transformation but formant structure changes according to the articulatory reorganization.}
  \label{fig:Figure4}
\end{figure}
The tuning of the six pair clusters of /b,d,g/ follows the same principle, but there is a special grouping of articulators when /b/ is involved in the pair. In this case $S_c=\{1,2,3,6\}$ and otherwise this is $S_c=\{1,2,3,4\}$. When involved, the /g/ is systematically palatal. The position of the /d/ slightly depends on the vowel as previously. Some languages systematically use stop pairs but if statistics are available for articulatory data (see \cite{Gafos2010}), there is no description of formant trajectories. This is thus difficult to evaluate our simulations with this criterion. We must admit this is challenging to form all pairs with all vowels in onset and coda position and some confusions persist after this tuning. 
\section{Conclusion about the title}
Finally, we simulate the verbal transformation from big.bi to bi.gbi of the title as an overview of the model properties. The graphs of /big/ and /bi/ are concatenated with $\delta_o=\delta_e=0.7$ to insert as in \cite{Gafos2002,Gafos2010} an audible schwa-like vocoid between the two consecutive consonants (Figure~\ref{fig:Figure3}~B). When $\delta_o=\delta_e$ tends to $1$ the graph reorganization can occur because two successive superimposed segments of duration $2*2T$ anchored to the same vowel $V_o=V_e=V$ can fuse into a single one of duration $3T$. At first, the gain of consonant clustering is a reduction of one period $T$. The graph complexity is corollary reduced with a decay of the number of nodes with the loss of $V$ as well as of the number of arcs (Figure~\ref{fig:Figure4}). Secondly, gestures are reorganized  because the selection process switches from two successive vectors $S_c=\{1,2,3,4\}$ for /g/ and $S_c=\{1,2,6\}$ for /b/ to a single vector $S_c=\{1,2,3,6\}$ for /gb/. This is clearly visible in Figure~\ref{fig:Figure4} where there is only one gesture for /gb/. Following Sato et al.'s claim \cite{Sato2006} of a  production constraint in perceptual verbal transformation, the syllabification has a preference for the most compact form as bi.gbi instead of big.bi. 

\bibliographystyle{IEEEtran}
\bibliography{mybib}

\end{document}